**Practical causal evaluation metrics for biological networks**

Noriaki Sato[1], Marco Scutari[2], Shuichi Kawano[3], Rui Yamaguchi[4,5], Seiya Imoto[1]

[1]Division of Health Medical Intelligence, Human Genome Center, the Institute of Medical Science, the University of Tokyo, Minato-ku, Tokyo, Japan.

[2]Istituto Dalle Molle di Studi sull'Intelligenza Artificiale (IDSIA), Lugano, Switzerland.

[3]Faculty of Mathematics, Kyushu University, Fukuoka, Japan.

[4]Division of Cancer System Biology, Aichi Cancer Center Research Institute, Nagoya, Japan.

[5]Division of Cancer Informatics, Nagoya University Graduate School of Medicine, Nagoya, Japan.

**Abstract**

Estimating causal networks from biological data is a critical step in systems biology. When evaluating the inferred network, assessing the networks based on their intervention effects is particularly important for downstream probabilistic reasoning and the identification of potential drug targets. In the context of gene regulatory network inference, biological databases are often used as reference sources. These databases typically describe relationships in a qualitative rather than quantitative manner. However, few evaluation metrics have been developed that take this qualitative nature into account. To address this, we developed a metric, the sign-augmented Structural Intervention Distance (sSID), and a weighted sSID that incorporates the net effects of the intervention. Through simulations and analyses of real transcriptomic datasets, we found that our proposed metrics could identify a different algorithm as optimal compared to conventional metrics, and the network selected by sSID had a superior performance in the classification task of clinical covariates using transcriptomic data. This suggests that sSID can distinguish networks that are structurally correct but functionally incorrect, highlighting its potential as a more biologically meaningful and practical evaluation metric.

**Introduction**

High-throughput omics have accelerated biological discovery through causal network inference.[1,2] Causal discovery algorithms specifically accounting for the nature of single-cell transcriptomics data have been proposed,[3] along with a benchmarking framework.[4,5]

Evaluating the inferred directed networks before downstream analyses, such as probabilistic reasoning and application in therapeutic target detection, is crucial for establishing the reliability and accuracy of the insights we extract from them. In particular, we want causal inference to be accurate to quantify precisely how pharmacological interventions on specific molecules propagate along relevant pathways. For this purpose, we need metrics that capture differences in the interventional distributions induced by the do-operator relative to some baseline.[6]

The first proposed metric of this kind is the Structural Intervention Distance (SID), which counts the number of incorrect interventional adjustments between the predicted and true graphs.[7] Its computational cost and inconsistencies in how it scores causally equivalent networks have led to the development of the *gadjid* framework.[8] Both metrics are purely graphical: they do not take into account either the magnitude of the differences in the interventional distributions nor their cost in terms of treatment misspecification. These limitations have been highlighted in more recent works. Eigennmann et al. proposed a framework for assessing causal structure learning methods using risk estimation.[9] Continuous SID (contSID) uses a kernel-based representation of the distribution of edge weights to enable assessment in graphs with weighted edges.[10] Also, CausalBench proposed the use of the mean Wasserstein distance for the evaluation of gene regulatory networks inferred from single-cell perturbation data, in terms of the evaluation of biological networks.[5,11]

Biological databases such as Kyoto Encyclopedia of Genes and Genomes (KEGG), TRRUST, OmniPath, and Reactome, provide canonical baselines for many bioinformatics tasks, such as gene (transcriptomic) regulatory network and protein-protein interaction inferences.[12–16] To design causal evaluation metrics for such tasks, we must (1) incorporate the available information of the effect sizes associated with the edges in the network and (2) combine them using a loss function that aligns well with our goal of quantifying the total effect of pharmacological interventions. Unfortunately, most biological databases contain only qualitative information on each edge (such as "stimulation", "activation", or "inhibition" based on the curated literature) and do not provide interaction strengths.

In such settings, we can only define a fully correct graph as one that captures both structure and regulatory sign. A correct structure ensures that interventions propagate correctly; the regulatory sign ensures that upward and downward regulation are correctly represented at the therapeutic target. Deviations from either imply that the network is not a correct mechanistic representation of the molecular network, and their severity should be weighted in proportion to their impact on the evaluation of pharmacological interventions. Currently, no metric natively captures the semantics that dominate biological knowledge bases in this way. SID counts the incorrect pairs without accounting for edge properties; contSID requires effect sizes as edge weights, which the databases do not supply; and neither method accounts for weighting errors based on their impact on treatment design.

Thus, here we develop a causal network evaluation metric with an explicit sign test for the practical use in biological settings, yielding a metric that (a) builds on SID and related metrics, (b) incorporates the discrete {0, ±1} interaction annotations from biological databases, (c) combines them in a way that is informative to treatment design and (d) reports scores on the same scale as SID, facilitating drop-in replacement in the assessment of causal structure in biological network inference tasks.

**Methods**

*Preliminaries*

SID measures differences between DAGs by counting the number of differences in the causal inference statements they imply. Formally, $SID = \sum_{i \neq j} SID(i, j)$, where $SID(i, j)$ is an indicator variable that is equal to one if the node $i$ in a learned graph $G_p$ fails to identify the correct interventional effect to the node $j$ in the true graph $G$.

*Sign-based evaluation*

Let $G = (V, E)$ be a DAG on $p$ variables. The total effect of $X_i$ on $X_j$ can be written as:

$$T(G)_{i \to j} = \left[ \sum_{k=1}^{p-1} B^k \right]_{ij},$$

where $B$ is the weighted adjacency matrix derived from $G$, and each cell indicates edge weights.[17] Our modified SID, named *sign-augmented SID* (sSID), adds one rule to the original SID, that if SID declares $(i, j)$ correct and the sign of $T(G)_{i \to j}$ is known but differs from the $T(G_p)_{i \to j}$ of the estimated graph $G_p$, mark the pair as incorrect.

Formally, let

$$S(B) = sign\left( \sum_{k=1}^{p-1} B^k \right) \in \{-1, 0, NA, 1\}^{p \times p}.$$

We define

$$sSID(G, G_p) = \sum_{i \neq j} [(1 - \lambda) SID(i,j) + \lambda(S(B)_{i \to j} \in \{\pm 1\} \wedge S(B)_{i \to j} \neq S(B^\wedge)_{i \to j})],$$

where $S(B)_{i \to j}$ indicates the sign of the total effect from the node $i$ to $j$, and $B^\wedge$ is the weighted adjacency matrix derived from $G_p$. If some path exist but none are fully annotated, we set $S(B)_{i \to j} = NA$. We additionally introduce a parameter $\lambda$ ($0<\lambda<1$) that controls the degree of weighing the SID and sign errors, which is set to 0.5 by default. Thus, sSID incorporates wrong paths as well as wrong regulatory signs.

An example of a transcriptional interaction network of the TGF pathway supported by the existing literature (73 variables and 146 edges) is shown in Figure 1A.[18,19] An illustrative example in the network consisting of three nodes is shown in Figure 1B. Biologically, if gene A encodes a transcription factor that represses gene B, and gene B in turn activates gene C, the pathway that over-expresses gene A will ultimately downregulate gene C, because the negative and positive regulations multiply to a net negative effect (Figure 1B, left, reference). Conversely, if both edges are activating, the signal remains positive (Figure 1B, center, denoted as $G_1$). As the topological structure of the reference and $G_1$ is identical, SID regards the pair $(A, C)$ as correct, and structural hamming distance (SHD) also reports no difference because it only considers edge presence and orientation. The contSID formulation is well-suited for comparing continuous interventional effect estimates, but when the reference labels are discrete sign annotations, the method cannot directly leverage the sign information.

By contrast, sSID evaluates not only the correctness of the structural relationships but also the agreement of the total effect's sign, and thus correctly identifies this pair as erroneous. From a biological standpoint, this distinction is critical because the predicted regulatory outcome of over-expressing A is reversed compared to the actual outcome, potentially leading to opposite conclusions in experimental interpretation. Of note, sSID can correctly capture errors in direction (Figure 1B, right, denoted as $G_2$), as do SID and SHD. Supplementary Table S1 illustrates each metric D when the reference, $G_1$, and $G_2$ are compared. The procedure for comparing the reference and $G_2$ is described in detail in Supplementary Text S1.

We also consider the absolute difference in the average total effect between nodes $X_i$ and $X_j$. We define $P_{ij}^{(k)}$ as the set of fully annotated paths, and $n_{ij}$ as the number of path lengths containing at least one fully annotated path from $i$ to $j$. $s_{ij}^{(k)}$ is the average net effect over all fully annotated paths of length $k$ from $i$ to $j$. We define the averaged total effect as:

$$A_{ij} = \frac{1}{n_{ij}} \sum_{k=1}^{p-1} s_{ij}^{(k)} \times 1(P_{ij}^{(k)} > 0)$$

Paths that contain one or more edges with unknown sign (NA) are excluded from the calculation, and if all possible paths from $i$ to $j$ contain an unknown sign, $A_{ij}$ is set to NA.

Given a reference graph $G$ and an estimated graph $G_p$, we quantify their discrepancy on a node pair by the absolute polarity difference computed only for pairs where both are defined:

$$\Delta_{ij} = \left| A_{ijG} - A_{ijG_p} \right|$$

The *weighted sign-augmented SID* (wsSID) aggregates these per-pair discrepancies across $i \neq j$, thereby capturing how far the estimated net effect deviates from the reference, not just whether the sign flips:

$$wsSID(G, G_p) = \sum_{i \neq j} [(1 - \lambda)SID(i, j) + \lambda \Delta_{ij}].$$

This metric can be interpreted as the net regulatory polarity, averaged across all possible regulatory edges between genes. For example, if gene A encodes a transcription factor that represses gene B, and gene B activates gene C, the only fully annotated path A→B→C has a negative sign product, yielding a negative $A_{AC}$. If additional annotated paths existed, their sign products would be incorporated into the average, reflecting the overall directionality of regulation. This approach captures aggregate polarity across all path lengths while ensuring that uncertain or partially annotated paths do not affect the estimate. This is suitable for applications where the degree of up- or down-regulation matters (e.g., partial cancellation among parallel cascades), complementing binary, sign-only criteria.

*Simulation studies*

To evaluate how each distance metric differs based on distinct types of inference error, we randomly generated DAGs with size $P = 20$ and an edge probability of 0.25. Every present edge was assigned an activation sign +1 or an inhibition sign −1 with equal probability. From that reference graph, we generated three perturbed variants, each constructed to isolate a single error mechanism. In the "extra-edge" case, we sequentially inserted a randomly chosen proportion of currently absent edges (0-30%). Directions were sampled so that half of the new arcs pointed towards existing nodes, thereby producing additional parents. For the "sign-flip" case, we retained the exact structure of the reference DAG but reversed the sign of a specified fraction of activation/inhibition edges (0-50%), leaving unknown-sign edges unchanged. In the "edge-drop" case, we removed a given proportion of true edges (0-30%) while preserving the remainder of the topology.

Subsequently, we generated random graphs based on the Barabási-Albert model with size $P = 10, 30, 50,$ and $70$, and an expected number of edges of 3. These numbers represent the typical number of genes in a biological pathway in real-world applications described in the following section. For each edge, we first sampled a sign uniformly at random from $\{-1,+1\}$. Subsequently, we drew the absolute magnitude from an exponential distribution with rate parameter $1/0.08$. The final edge weight was given by the product of the random sign and the exponential magnitude. Treating this weighted matrix $B$ as the structural equation model, we simulated $n=300$ observations from the linear models using the rmvDAG function in pcalg.

Using the same data set, we inferred causal graphs with the algorithms described in the following section. For each learned graph, we reconstructed a coefficient matrix by regressing every child on its parent set and then evaluated the metrics against the original DAG. This procedure was replicated ten times per error condition, allowing us to compare how reliably each metric discriminates structure, adjustment validity, and sign accuracy across the tested algorithms.

*Validation of wsSID*

To isolate the effect of polarity composition while holding graph structure fixed, we constructed signed DAGs in which a single source-target pair $(i, j)$ is connected by $K$ parallel paths of equal length $L = 2$. Each path consists of $L - 1$ intermediate nodes arranged in series from $i$ to $j$. Edge directions and presence are identical in the reference and estimated graphs, ensuring that SHD and SID remain zero. Edge weights are restricted to $\pm 1$, and no edges are annotated as NA. In the reference graph, all $K$ paths from $i$ to $j$ are assigned positive polarity, so the total effect sign along any path equals $+1$. In the estimated graph, we vary the fraction of negative paths by flipping the sign of the first edge on $r \in \{0, ..., K\}$ of the $K$ parallel paths, leaving all other edges unchanged; the structural adjacency thus remains identical, but the composition of positive versus negative paths changes smoothly with the ratio $r/K$. For each setting, we compute two quantities on the pair $(i, j)$. First, we compute the average polarity $A_{ij}$. Second, we compute sSID on $(i, j)$ using the sign of the total effect implied by all fully annotated paths: sSID returns 0 when the estimated polarity matches the reference and 1 when it differs.

*Comparison based on actual biological databases*

We first constructed a set of DAGs from the pathways for the transcriptomic gene regulatory interactions stored in the OmniPath databases.[14,18] We extracted the interactions whose source or target gene is in the candidate signaling pathway based on SignaLink pathway annotation using the R package OmniPathR.[19] The edge annotation of "stimulation" was coded as 1 and the "inhibition" as -1. The feedback arc set was identified with the igraph function, and the corresponding edges were removed from the graph. The resulting DAGs derived from the pathways were used as the reference in the evaluation. The DAGs with fewer than 10 nodes (genes) were not evaluated.

As the input for structure learning, the gene expression data from lung adenocarcinoma (LUAD) in The Cancer Genome Atlas (TCGA) were used.[20] The $\log_2(x+1)$ transformed and RSEM (RNA-Seq by Expectation Maximization) normalized count was downloaded from UCSC Xena browser.[21]

*Prediction of tumor stage based on inferred DAGs*

Furthermore, we conducted an experiment to predict categories of TCGA clinical covariates using DAGs inferred by multiple methods. Specifically, we inferred DAGs using the genes included in the KEGG PATHWAY "Prostate cancer" (hsa05215) from the TCGA-PRAD dataset. We then generated features based on the networks selected as best by SID, sSID, and wsSID, and used these features to predict the high stage (>=3) with a random forest classifier.[22] Finally, we compared the networks' performance using the area under the precision-recall curve (AUPRC). Here, the high stage is defined as T stage $\geq$ T3 (preferring pathologic T over clinical T when available) or if pathologic nodes are positive (N1). We construct DAG-based features directly from the DAGs by modeling each node's expression as a linear function of its parents, and using the predicted expression values as the feature (using the predict method in bnlearn).

*Structure learning algorithms*

We tested five representative algorithms: Max-min hill climbing (MMHC) with the alpha threshold of 0.05 and 0.01 in the restricting phase, extremely greedy equivalence search (XGES), GES, ICA-LiNGAM, and DirectLiNGAM.[23–26] GES and ICA-LiNGAM are implemented in the R package pcalg and are used with the default hyperparameters. DirectLiNGAM and MMHC are implemented in the R package bnlearn.[27,28] The regression coefficients, indicating the edge weights used for the evaluation for each node, were learned by ordinary least squares for DirectLiNGAM, MMHC, XGES, and GES using the bn.fit function in the bnlearn.

*Compared metrics*

For each simulation setting, we regenerated 10 independent replicates and computed the SHD, the original SID, and the sSID against the reference graph. We used the SHD in the R package bnlearn, and the SID implemented in the CRAN package of the same name. For contSID, currently, no implementation is available, and we do not include the metrics in the comparison. The implementation of sSID is available at https://github.com/norakis/sSID. The R version 4.4.1 was used in the manuscript. The visualization and graph manipulation were performed using the R libraries ggraph, tidygraph, and igraph.[29]

**Results**

*Only sSID and wsSID captured the sign error correctly*

We first performed simulation experiments modifying the structure of the true DAG (Figure 1C). When adding edges, sSID and wsSID increase compared to the original SID because the added edges carry random signs that impact downstream effects. When dropping edges, sSID rises almost in parallel with SID, and wsSID tracks the same trend. When flipping the signs of the edges, SHD and SID remain at zero (structure and adjustment sets are intact). In contrast, sSID increases almost linearly with the flip rate, proving that the new sign test detects sign errors invisible to existing metrics, as expected.

*wsSID captured the degree of regulatory differences*

The reference graph used for wsSID validation is shown in Figure 1D. Sweeping the negative-path ratio from $r/K=0$ to 1 produces a continuous transition in the average polarity $-1$, with a precise cancellation point at $r/K=0.5$ where the contributions of positive and negative paths balance and $A_{ij}=0$ (Figure 1D, right). This behavior captures the gradual growth of net regulatory influence as increasingly many opposing paths are introduced. In contrast, sSID remains zero throughout the entire regime in which the majority polarity agrees with the reference and then switches discretely to one once the estimated total-effect sign flips from positive to negative. Consequently, sSID highlights the biologically critical threshold at which the functional conclusion reverses (activation versus repression), whereas the average polarity quantifies how close the system is to that reversal by revealing the degree of cancellation among parallel routes. As intended, SHD and SID remain at zero across the sweep, confirming that the observed differences are driven by polarity composition under identical topology.

*Different algorithms were optimal in the simulated dataset*

Next, we inferred networks from observational data simulated from a DAG under the Barabási-Albert model using multiple algorithms, and compared their performance using the metrics (Table 1). Across different values of *P*, the algorithms favored by SID, sSID, and wsSID were not always consistent. In both $P = 30$ and P = 70, all metrics identified XGES as the best-performing method. In *P* = 10, while SID favored XGES and SHD favored GES, sSID and wsSID identified DirectLiNGAM as the best. In *P* = 50, SHD, SID, and wsSID selected XGES, while sSID favored MMHC (α=0.01). The reason for the discrepancy between SID and sSID could be that SID evaluates only structural reachability, whereas sSID additionally requires the correct sign of interventional effects. Methods that recover true interventional effects yet misestimate effect directions along direct or mediated paths perform worse on sSID. The discrepancy between sSID and wsSID here was that denser models (with higher edge counts) create more opportunities to correctly match effect signs, whereas sparser models (with fewer FP edges) avoid high-weight wrong-sign paths. These findings demonstrate that the algorithm considered optimal can differ depending on whether the evaluation metric is based on structural accuracy or incorporates sign information (sSID and wsSID). However, the results may vary depending on the network simulation model and the number of variables.

*Different algorithms were optimal in networks inferred from the actual biological dataset*

Furthermore, using DAGs derived from actual biological datasets as references, we inferred networks based on gene expression data from TCGA-LUAD and compared the resulting metrics. The constructed DAG contains 20-193 variables with 31-472 edges. The results are shown in Supplementary Table S2. Consistent with the simulation study, the algorithm that appeared best depended on the metric assessed across several pathways: LiNGAM attained the lowest SID, and MMHC (α=0.01) attained the lowest SHD, while GES minimised sSID in the TGF pathway.

This disparity confirms that, in the actual biological data, structural fit, intervention validity, and sign properties capture independent aspects of causal structure evaluation, as in the simulated data. Therefore, relying on a single metric would misrank the competing algorithms and could mask functionally critical sign errors.

*The algorithm selected by sSID had better performance in classification*

To further demonstrate the consequences of these discrepancies, we conducted a classification experiment based on network-predicted gene expression data. The reference DAG extracted from Dorothea is depicted in Supplementary Figure S1. Among the networks inferred from the TCGA-PRAD data, sSID and wsSID identified the GES-inferred network as the best, whereas SID and SHD selected the MMHC (α= 0.01)-inferred network as the best network. Using gene expression values predicted from these networks, we performed a random forest-based classification task to predict samples with high tumor stage. In this task, the network inferred by GES showed the best performance (right panel, Figure 1E). All metric results are summarized in Supplementary Table S3. These findings suggest that the proposed sSID is useful for selecting networks for biologically meaningful analysis tasks.

**Discussion**

As a limitation, applying sSID to networks learned by methods without full identifiability raises issues of Markov equivalence. Because sSID is defined on fully oriented, signed DAGs, scoring a single representative can conflate unresolved orientation ambiguity with true error. Accordingly, sSID can detect sign errors only where the sign of the total effect is identifiable. Also, the proposed sSID metric is applicable only when linear or monotonic relationships between nodes are assumed, because edge signs and coefficients can be consistently defined. In general, in nonlinear or nonparametric DAG models, sSID cannot be computed directly, whereas SHD and SID remain applicable.

In this paper, we presented the causal-graph distance tailored to biological pathway annotations. By integrating edge sign or polarity into the SID framework, sSID and wsSID distinguish graphs that are structurally correct yet functionally incorrect, which is an essential capability for reliable downstream biological interpretation.

**Resource availability**

Lead contact: Noriaki Sato, nori@hgc.jp

Materials availability: No materials were generated in the study.

Data and code availability: The code of the implementation is available at https://github.com/noriakis/sSID. Note that currently the repository is private and the source codes are attached to the submission.

**Acknowledgement**

The study is partially supported by Takeda Science Foundation and JSPS KAKENHI Grant Number 25K21331.

**Author contributions**

N.S. conceived and designed the study, developed the method, performed the analyses, and drafted the manuscript. M.S. contributed to the theoretical framework and algorithmic design. S.K. and R.Y. provided statistical guidance and contributed to methodological refinement. S.I. supervised the overall project and contributed to the study design, interpretation, and manuscript revision. All authors read and approved the final manuscript.

**Declaration of Interests**

The authors declare no competing interests.

**Supplemental information**

Supplementary Text S1. The detailed calculation procedure of sSID.

Supplementary Figure S1. The reference transcriptomic interaction network related to prostate cancer.

Supplementary Table S1. Illustrative example of sSID.

Supplementary Table S2. The metrics summary for biological pathways.

Supplementary Table S3. The classification performance and network assessment metrics.

**Figure titles and legends**

**Figure 1. Sign-augmented SID.** (A) An example of a transcriptional interaction network (TGF pathway) represented as a directed acyclic graph extracted from Dorothea database. The red arc indicates stimulation and the blue indicates inhibition, supported by the curated literature. The node size indicates the centraliy degree. (B) An illustrative example of proposed metrics. The reference graph (left), G_1 (center), and G_2 (right). (C) Simulation of modifications in the true graph. When randomly flipping the signs of edges, only sSID is sensitive to the modifications (center plot, P=20). The lines of SHD and SID are perfectly overlapped in the center plot. (D) Validation of weighted sSID (wsSID). (left) The reference graph used for the validation. Parallel paths of length 2 were assigned the label of signs. (right) Line plot depicting the sSID (dashed) and wsSID (bold) across the different fractions of negative paths (right). (E) The classification performance assessment based on the algorithms chosen by SID, sSID, SHD, and wsSID. The reference network used for the assessment and the precision-recall curve (PRC) are shown. The value inside the brackets indicates the area under PRC.

**Tables**

**Table 1. Comparison between structure learning algorithms**

| $P$ | Algorithm | SID_mean | sSID_mean | SHD_mean | wsSID_mean |
|---|---|---|---|---|---|
| 10 | DirectLiNGAM | 64.6 | **35.3** | 20.4 | **36.31527778** |
| 10 | GES | 63.6 | 36.95 | **17** | 37.07944444 |
| 10 | ICA-LiNGAM | 66.1 | 38.25 | 19 | 37.70055556 |
| 10 | MMHC_0.01 | 64.7 | 37.4 | 17.8 | 36.72361111 |
| 10 | MMHC_0.05 | 63.5 | 36.6 | 17.5 | 36.64111111 |
| 10 | XGES | **63** | 37.5 | 18 | 36.65083333 |
| 30 | DirectLiNGAM | 592.5 | 311.1 | 103.7 | 328.0895761 |
| 30 | GES | 582.7 | 315.7 | 59.6 | 317.5857381 |
| 30 | ICA-LiNGAM | 555.9 | 312.65 | 62.3 | 308.4451508 |
| 30 | MMHC_0.01 | 585.3 | 319.85 | 59.6 | 317.1941151 |
| 30 | MMHC_0.05 | 586.1 | 318.35 | 59.6 | 318.6210198 |
| 30 | XGES | **549.2** | **308.15** | **58.3** | **304.1453333** |
| 50 | DirectLiNGAM | 1618.8 | 849.8 | 236 | 899.4326098 |
| 50 | GES | 1566.1 | 843 | 113.3 | 857.6330946 |
| 50 | ICA-LiNGAM | 1440.9 | 806.85 | 109 | 797.2859394 |
| 50 | MMHC_0.01 | 1444.6 | **796.15** | 102.1 | 789.7610564 |
| 50 | MMHC_0.05 | 1541.1 | 830.8 | 111 | 844.6246744 |
| 50 | XGES | **1439.8** | 798.9 | **96.9** | **789.4947804** |
| 70 | DirectLiNGAM | 3207.9 | 1650.3 | 440.3 | 1735.293571 |
| 70 | GES | 2802.7 | 1496.05 | 181.2 | 1566.453545 |
| 70 | ICA-LiNGAM | 2698.9 | 1471.1 | 164.1 | 1468.101292 |
| 70 | MMHC_0.01 | 2675.3 | 1448.5 | 151.1 | 1445.227809 |
| 70 | MMHC_0.05 | 2735.1 | 1468.35 | 169.8 | 1502.151821 |
| 70 | XGES | **2577.1** | **1421.5** | **137.8** | **1408.12804** |

The best-performing algorithms are shown in bold for each variable size (*P*).

**References**

1. Zhang, Y., Li, Q., Wang, J., Chang, X., Chen, L., and Liu, X. (2025). Causal network inference based on cross-validation predictability. Commun. Phys. *8*, 173. https://doi.org/10.1038/s42005-025-02091-4.

2. Sachs, K., Perez, O., Pe'er, D., Lauffenburger, D.A., and Nolan, G.P. (2005). Causal Protein-Signaling Networks Derived from Multiparameter Single-Cell Data. Science *308*, 523–529. https://doi.org/10.1126/science.1105809.

3. Choi, J., and Ni, Y. (2023). Model-based Causal Discovery for Zero-Inflated Count Data. J. Mach. Learn. Res. *24*, 1–32.

4. Pratapa, A., Jalihal, A.P., Law, J.N., Bharadwaj, A., and Murali, T.M. (2020). Benchmarking algorithms for gene regulatory network inference from single-cell transcriptomic data. Nat. Methods *17*, 147–154. https://doi.org/10.1038/s41592-019-0690-6.

5. Chevalley, M., Roohani, Y., Mehrjou, A., Leskovec, J., and Schwab, P. (2023). CausalBench: A Large-scale Benchmark for Network Inference from Single-cell Perturbation Data. Preprint at arXiv, https://doi.org/10.48550/arXiv.2210.17283 https://doi.org/10.48550/arXiv.2210.17283.

6. Pearl, J. (1995). A Causal Calculus for Statistical Research. In Pre-proceedings of the Fifth International Workshop on Artificial Intelligence and Statistics (PMLR), pp. 430–449.

7. Peters, J., and Bühlmann, P. (2015). Structural Intervention Distance for Evaluating Causal Graphs. Neural Comput. *27*, 771–799. https://doi.org/10.1162/NECO_a_00708.

8. Henckel, L., Würtzen, T., and Weichwald, S. (2024). Adjustment identification distance: a gadjid for causal structure learning. In Proceedings of the Fortieth Conference on Uncertainty in Artificial Intelligence UAI '24. (JMLR.org), pp. 1569–1598.

9. Eigenmann, M., Mukherjee, S., and Maathuis, M. (2020). Evaluation of Causal Structure Learning Algorithms via Risk Estimation. In Proceedings of the 36th Conference on Uncertainty in Artificial Intelligence (UAI) (PMLR), pp. 151–160.

10. Dhanakshirur, M., Laumann, F., Park, J., and Barahona, M. (2025). A Continuous Structural Intervention Distance to Compare Causal Graphs. In Causal Inference, X.-H. Zhou and J. Jia, eds. (Springer Nature), pp. 25–40. https://doi.org/10.1007/978-981-97-7812-6_3.

11. Chevalley, M., Sackett-Sanders, J., Roohani, Y., Notin, P., Bakulin, A., Brzezinski, D., Deng, K., Guan, Y., Hong, J., Ibrahim, M., et al. (2025). The CausalBench challenge: A machine learning contest for gene network inference from single-cell perturbation data. Preprint at arXiv, https://doi.org/10.48550/arXiv.2308.15395 https://doi.org/10.48550/arXiv.2308.15395.

12. Kanehisa, M., and Goto, S. (2000). KEGG: kyoto encyclopedia of genes and genomes. Nucleic Acids Res. *28*, 27–30. https://doi.org/10.1093/nar/28.1.27.

13. Jassal, B., Matthews, L., Viteri, G., Gong, C., Lorente, P., Fabregat, A., Sidiropoulos, K., Cook, J., Gillespie, M., Haw, R., et al. (2020). The reactome pathway knowledgebase. Nucleic Acids Res *48*, D498–D503. https://doi.org/10.1093/nar/gkz1031.

14. Türei, D., Korcsmáros, T., and Saez-Rodriguez, J. (2016). OmniPath: guidelines and gateway for literature-curated signaling pathway resources. Nat. Methods *13*,

966–967. https://doi.org/10.1038/nmeth.4077.

15. Han, H., Cho, J.-W., Lee, S., Yun, A., Kim, H., Bae, D., Yang, S., Kim, C.Y., Lee, M., Kim, E., et al. (2018). TRRUST v2: an expanded reference database of human and mouse transcriptional regulatory interactions. Nucleic Acids Res. *46*, D380–D386. https://doi.org/10.1093/nar/gkx1013.

16. Zhang, Y., Chang, X., and Liu, X. (2021). Inference of gene regulatory networks using pseudo-time series data. Bioinformatics *37*, 2423–2431. https://doi.org/10.1093/bioinformatics/btab099.

17. Bollen, K.A. (1987). Total, Direct, and Indirect Effects in Structural Equation Models. Sociol. Methodol. *17*, 37–69. https://doi.org/10.2307/271028.

18. Garcia-Alonso, L., Holland, C.H., Ibrahim, M.M., Turei, D., and Saez-Rodriguez, J. (2019). Benchmark and integration of resources for the estimation of human transcription factor activities. Genome Res. *29*, 1363–1375. https://doi.org/10.1101/gr.240663.118.

19. Fazekas, D., Koltai, M., Türei, D., Módos, D., Pálfy, M., Dúl, Z., Zsákai, L., Szalay-Bekő, M., Lenti, K., Farkas, I.J., et al. (2013). SignaLink 2 – a signaling pathway resource with multi-layered regulatory networks. BMC Syst. Biol. *7*, 7. https://doi.org/10.1186/1752-0509-7-7.

20. Cancer Genome Atlas Research Network, Weinstein, J.N., Collisson, E.A., Mills, G.B., Shaw, K.R.M., Ozenberger, B.A., Ellrott, K., Shmulevich, I., Sander, C., and Stuart, J.M. (2013). The Cancer Genome Atlas Pan-Cancer analysis project. Nat Genet *45*, 1113–1120. https://doi.org/10.1038/ng.2764.

21. Goldman, M.J., Craft, B., Hastie, M., Repečka, K., McDade, F., Kamath, A., Banerjee, A., Luo, Y., Rogers, D., Brooks, A.N., et al. (2020). Visualizing and interpreting cancer genomics data via the Xena platform. Nat Biotechnol *38*, 675–678. https://doi.org/10.1038/s41587-020-0546-8.

22. Wright, M.N., and Ziegler, A. (2017). ranger: A Fast Implementation of Random Forests for High Dimensional Data in C++ and R. J. Stat. Softw. *77*, 1–17. https://doi.org/10.18637/jss.v077.i01.

23. Shimizu, S., Inazumi, T., Sogawa, Y., Hyvärinen, A., Kawahara, Y., Washio, T., Hoyer, P.O., and Bollen, K. (2011). DirectLiNGAM: A Direct Method for Learning a Linear Non-Gaussian Structural Equation Model. J. Mach. Learn. Res. *12*, 1225–1248.

24. Shimizu, S., Hoyer, P.O., Hyvä, A., rinen, and Kerminen, A. (2006). A Linear Non-Gaussian Acyclic Model for Causal Discovery. J. Mach. Learn. Res. *7*, 2003–2030.

25. Nazaret, A., and Blei, D. (2024). Extremely greedy equivalence search. In Proceedings of the Fortieth Conference on Uncertainty in Artificial Intelligence UAI '24. (JMLR.org), pp. 2716–2745.

26. Tsamardinos, I., Brown, L.E., and Aliferis, C.F. (2006). The max-min hill-climbing Bayesian network structure learning algorithm. Mach. Learn. *65*, 31–78. https://doi.org/10.1007/s10994-006-6889-7.

27. Scutari, M. (2010). Learning Bayesian Networks with the bnlearn R Package. J. Stat. Softw. Artic. *35*, 1–22. https://doi.org/10.18637/jss.v035.i03.

28. Markus Kalisch, Martin Mächler, Diego Colombo, Marloes H. Maathuis, and Peter Bühlmann (2012). Causal Inference Using Graphical Models with the R Package pcalg. J. Stat. Softw. *47*, 1–26. https://doi.org/10.18637/jss.v047.i11.

29. Csárdi, G., and Nepusz, T. (2006). The igraph software package for complex network research. InterJournal *Complex Systems*, 1695.

**Supplementary Table S1. Illustrative example of sSID.** Comparison of metrics between (reference, $G_1$) and (reference, $G_2$).

| Metrics | (Reference, $G_1$) | (Reference, $G_2$) |
|---|---|---|
| SHD | 0 | 2 |
| SID | 0 | 3 |
| sSID ($\lambda = 0.5$) | 1 | 2.5 |

**Supplementary Text S1. The detailed calculation procedure of sSID.**
Let $B(R)$ and $B(G_2)$ denotes the weighted adjacency matrix of reference and $G_2$ graph:
$$B(R) = [0\ -1\ 0\ 0\ 0\ 1\ 0\ 0\ 0], B(G_2) = [0\ 1\ 0\ 0\ 0\ 0\ 0\ 1\ 0].$$
The total effect of reference and G2 can be computed as:
$$T(R) = [0\ -1\ -1\ 0\ 0\ 1\ 0\ 0\ 0] \text{ and } T(G_2) = [0\ 1\ 0\ 0\ 0\ 0\ 0\ 1\ 0]$$
Original SID between reference and $G_2$ is 3, and the incorrect matrix, indicating $SID(i, j)$ per cell, of SID is:
$$[0\ 0\ 0\ 0\ 0\ 1\ 1\ 1\ 0].$$
The additionally flagged matrix of different signs of the net effect where the total effect differs is:
$$[0\ 1\ 1\ 0\ 0\ 1\ 1\ 1\ 0].$$
Thus, for λ=0.5, we compute sSID as 3×0.5+2×0.5=2.5 and for λ=0.8, 3×0.2+2×0.8=3.

Supplementary Table S2. The metrics summary for biological pathways.

| Algorithm | SID | sSID | SHD | wsSID | edges | TP | FP | FN | Pathway | P |
|---|---|---|---|---|---|---|---|---|---|---|
| DirectLiNGAM | 2928 | 1496 | 769 | 1833.45811 | 625 | 16 | 609 | 177 | JAK/STAT | 89 |
| GES | 3069 | 1567 | 963 | 1835.59529 | 837 | 24 | 813 | 169 | JAK/STAT | 89 |
| ICA-LiNGAM | 2747 | 1434 | 361 | 1633.851009 | 190 | 9 | 181 | 184 | JAK/STAT | 89 |
| MMHC_0.01 | 2590 | 1361.5 | 342 | 1770.978322 | 171 | 11 | 160 | 182 | JAK/STAT | 89 |
| MMHC_0.05 | 2639 | 1380 | 376 | 1836.536595 | 206 | 11 | 195 | 182 | JAK/STAT | 89 |
| XGES | 2921 | 1503 | 637 | 1742.274472 | 488 | 15 | 473 | 178 | JAK/STAT | 89 |
| DirectLiNGAM | 242 | 129 | 99 | 148.9998481 | 87 | 7 | 80 | 26 | Notch | 22 |
| GES | 268 | 142 | 99 | 157.6428138 | 83 | 5 | 78 | 28 | Notch | 22 |
| ICA-LiNGAM | 234 | 135 | 74 | 155.8905556 | 49 | 2 | 47 | 31 | Notch | 22 |
| MMHC_0.01 | 292 | 151.5 | 58 | 171.0333333 | 33 | 4 | 29 | 29 | Notch | 22 |
| MMHC_0.05 | 289 | 152 | 66 | 173.6533333 | 43 | 4 | 39 | 29 | Notch | 22 |
| XGES | 275 | 145 | 77 | 163.7217207 | 53 | 4 | 49 | 29 | Notch | 22 |
| DirectLiNGAM | 8687 | 4362 | 1185 | 4537.623041 | 1016 | 23 | 993 | 214 | uclear hormone recepto | 114 |
| GES | 8478 | 4270 | 1285 | 4465.868365 | 1117 | 20 | 1097 | 217 | uclear hormone recepto | 114 |
| ICA-LiNGAM | 8681 | 4402 | 544 | 4591.788103 | 329 | 10 | 319 | 227 | uclear hormone recepto | 114 |
| MMHC_0.01 | 8353 | 4306 | 466 | 4637.753743 | 245 | 5 | 240 | 232 | uclear hormone recepto | 114 |
| MMHC_0.05 | 8864 | 4501.5 | 527 | 4837.471031 | 310 | 7 | 303 | 230 | uclear hormone recepto | 114 |
| XGES | 8533 | 4330.5 | 855 | 4560.391909 | 662 | 17 | 645 | 220 | uclear hormone recepto | 114 |
| DirectLiNGAM | 29923 | 15041.5 | 2324 | 15363.94836 | 1964 | 47 | 1917 | 425 | Receptor tyrosine kinase | 193 |
| GES | 30233 | 15137.5 | 3004 | 15308.67041 | 2669 | 48 | 2621 | 424 | Receptor tyrosine kinase | 193 |
| ICA-LiNGAM | 29068 | 14813 | 1606 | 15259.75487 | 1201 | 24 | 1177 | 448 | Receptor tyrosine kinase | 193 |
| MMHC_0.01 | 29350 | 14933 | 827 | 15543.71883 | 383 | 10 | 373 | 462 | Receptor tyrosine kinase | 193 |
| MMHC_0.05 | 29230 | 14848 | 909 | 15305.22205 | 469 | 10 | 459 | 462 | Receptor tyrosine kinase | 193 |
| XGES | 31015 | 15534 | 1912 | 15705.90342 | 1521 | 29 | 1492 | 443 | Receptor tyrosine kinase | 193 |
| DirectLiNGAM | 11126 | 5580 | 1039 | 5644.093006 | 847 | 20 | 827 | 234 | T-cell receptor | 113 |
| GES | 11024 | 5521.5 | 1384 | 5542.842387 | 1219 | 37 | 1182 | 217 | T-cell receptor | 113 |
| ICA-LiNGAM | 11031 | 5614 | 610 | 5670.279272 | 382 | 11 | 371 | 243 | T-cell receptor | 113 |
| MMHC_0.01 | 11213 | 5667 | 465 | 5816.812817 | 225 | 5 | 220 | 249 | T-cell receptor | 113 |
| MMHC_0.05 | 11166 | 5649 | 520 | 5777.387444 | 285 | 6 | 279 | 248 | T-cell receptor | 113 |
| XGES | 11148 | 5600 | 928 | 5646.487319 | 732 | 23 | 709 | 231 | T-cell receptor | 113 |
| DirectLiNGAM | 4019 | 2017 | 702 | 2075.085493 | 615 | 20 | 595 | 126 | TGF | 73 |
| GES | 3697 | 1857 | 733 | 1911.873384 | 642 | 15 | 627 | 131 | TGF | 73 |
| ICA-LiNGAM | 3602 | 1928 | 296 | 1986.968092 | 165 | 6 | 159 | 140 | TGF | 73 |
| MMHC_0.01 | 3753 | 1977 | 288 | 2099.064013 | 157 | 5 | 152 | 141 | TGF | 73 |
| MMHC_0.05 | 3974 | 2035 | 321 | 2162.479108 | 194 | 8 | 186 | 138 | TGF | 73 |
| XGES | 3978 | 2005 | 507 | 2083.090739 | 390 | 8 | 382 | 138 | TGF | 73 |
| DirectLiNGAM | 243 | 129.5 | 107 | 152.6374149 | 89 | 10 | 79 | 27 | TNF pathway | 22 |
| GES | 249 | 130.5 | 108 | 151.0565498 | 90 | 4 | 86 | 33 | TNF pathway | 22 |
| ICA-LiNGAM | 259 | 138 | 83 | 167.1508854 | 56 | 5 | 51 | 32 | TNF pathway | 22 |
| MMHC_0.01 | 284 | 149 | 69 | 170.2822376 | 41 | 2 | 39 | 35 | TNF pathway | 22 |
| MMHC_0.05 | 248 | 137 | 81 | 165.607128 | 54 | 0 | 54 | 37 | TNF pathway | 22 |
| XGES | 353 | 181 | 86 | 187.9886029 | 64 | 5 | 59 | 32 | TNF pathway | 22 |
| DirectLiNGAM | 186 | 98 | 99 | 113.9621678 | 82 | 2 | 80 | 29 | Toll-like receptor | 20 |
| GES | 176 | 97.5 | 85 | 115.2461049 | 74 | 7 | 67 | 24 | Toll-like receptor | 20 |
| ICA-LiNGAM | 129 | 74.5 | 76 | 105.4057926 | 53 | 2 | 51 | 29 | Toll-like receptor | 20 |
| MMHC_0.01 | 167 | 95.5 | 55 | 128.4583333 | 32 | 3 | 29 | 28 | Toll-like receptor | 20 |
| MMHC_0.05 | 178 | 100 | 61 | 125.0412698 | 37 | 2 | 35 | 29 | Toll-like receptor | 20 |
| XGES | 210 | 112 | 69 | 127.4733135 | 49 | 4 | 45 | 27 | Toll-like receptor | 20 |
| DirectLiNGAM | 1801 | 920.5 | 446 | 1008.73704 | 383 | 12 | 371 | 83 | WNT | 55 |
| GES | 2164 | 1091 | 463 | 1119.327766 | 399 | 14 | 385 | 81 | WNT | 55 |
| ICA-LiNGAM | 2051 | 1064 | 203 | 1120.688426 | 122 | 4 | 118 | 91 | WNT | 55 |
| MMHC_0.01 | 2101 | 1083.5 | 197 | 1157.526987 | 113 | 5 | 108 | 90 | WNT | 55 |
| MMHC_0.05 | 2089 | 1081.5 | 222 | 1156.221929 | 140 | 6 | 134 | 89 | WNT | 55 |
| XGES | 2330 | 1170 | 332 | 1198.023798 | 257 | 9 | 248 | 86 | WNT | 55 |

| | auc | pr_auc | SID | sSID | wsSID | SHD |
|---|---|---|---|---|---|---|
| **Supplementary Table S3. The classification performance and network assessment metrics.** | | | | | | |
| ICA-LiN | 0.635007715 | 0.743129132 | 1781 | 936 | 956.2213644 | 234 |
| DirectLiN | 0.645079719 | 0.737829067 | 1813 | 939 | 983.2093751 | 328 |
| MMHC_ | 0.672955597 | 0.749043242 | 1734 | 916 | 985.3023989 | 243 |
| MMHC_ | 0.666115207 | 0.743755582 | 1710 | 922.5 | 1003.74742 | 221 |
| XGES | 0.674215669 | 0.7593331 | 1845 | 939.5 | 976.6934235 | 331 |
| GES | 0.683044745 | 0.763753843 | 1795 | 913.5 | 942.6702806 | 430 |
| random | 0.656874679 | 0.732141698 | 1716 | 926 | 950.2659155 | 246 |

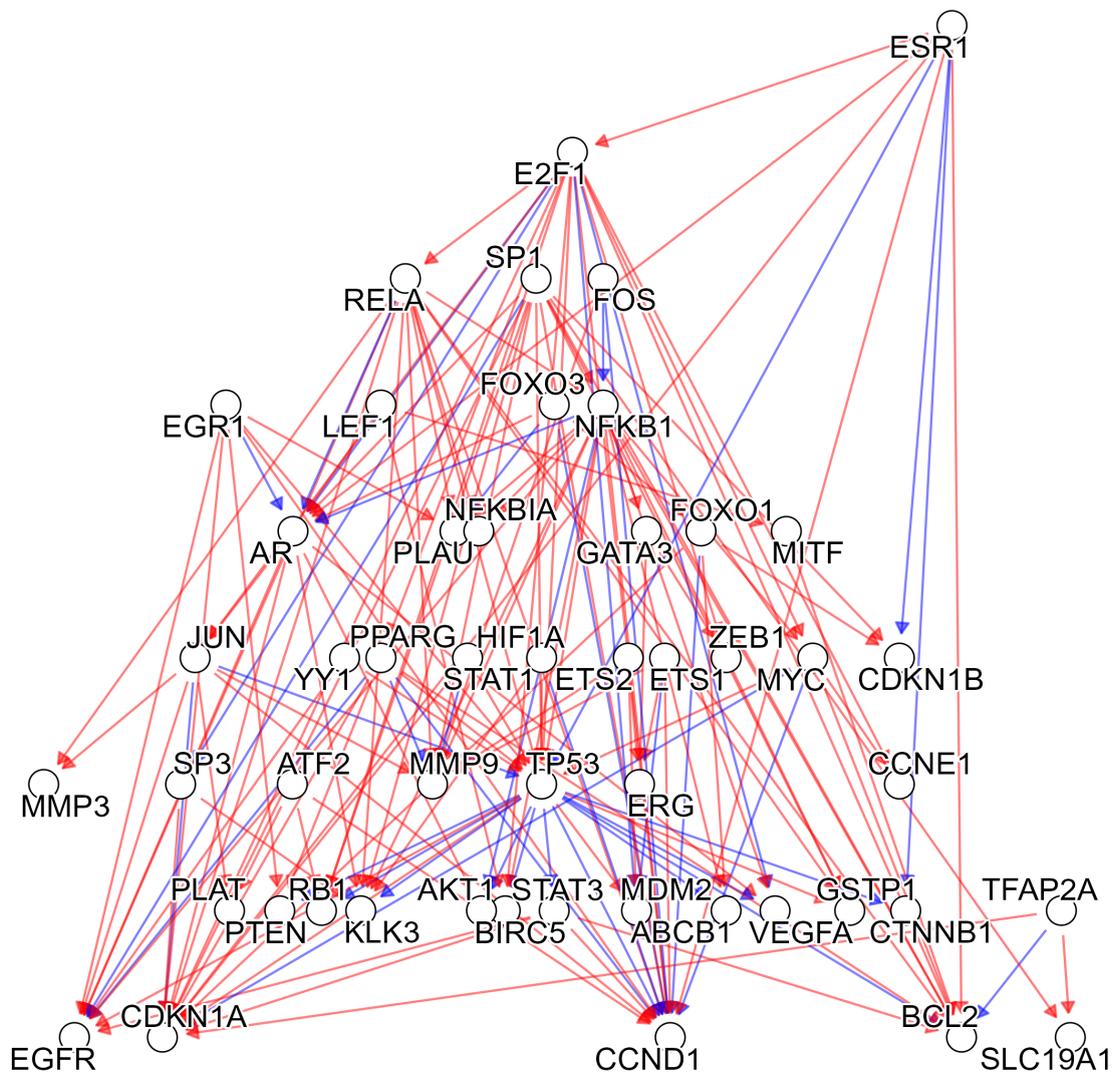

Supplementary Figure S1. The reference transcriptomic interaction network related to prostate cancer.